# Strain-induced enhancement of $T_c$ in infinite-layer $Pr_{0.8}Sr_{0.2}NiO_2$ films

Xiaolin Ren[1,2], Jiarui Li[3], Wei-Chih Chen[4], Qiang Gao[1], Joshua J. Sanchez[3], Jordyn Hales[4], Hailan Luo[1,2], Fanny Rodolakis[5], Jessica L. McChesney[5], Tao Xiang[1,6], Jiangping Hu[1,7], Fu-Chun Zhang[8], Riccardo Comin[3], Yao Wang[4,*], X. J. Zhou[1,2,6,7,*], and Zhihai Zhu[1,2,7,*]

[1]*Institute of Physics, National Laboratory for Condensed Matter Physics, Chinese Academy of Sciences, Beijing 100190, China*

[2]*School of Physical Sciences, University of Chinese Academy of Sciences, Beijing 100049, China*

[3]*Department of Physics, Massachusetts Institute of Technology, Cambridge, Massachusetts 02139, USA*

[4]*Department of Physics and Astronomy, Clemson University, Clemson, SC 29631, USA*

[5]*X-Ray Science Division, Advanced Photon Source, Argonne National Laboratory, Lemont, IL 60439, USA*

[6]*Beijing Academy of Quantum Information Sciences, Beijing 100193, China*

[7]*Songshan Lake Materials Laboratory, Dongguan 523808, China*

[8]*Kavli Institute for Theoretical Sciences and CAS Center for Topological Quantum Computation, University of Chinese Academy of Sciences，Beijing 100190, China*

(Dated: March 20, 2022)

---

[*] To whom correspondence should be addressed.

Emails: yaowang@g.clemson.edu, XJZhou@iphy.ac.cn, zzh@iphy.ac.cn

**The mechanism of unconventional superconductivity in correlated materials remains a great challenge in condensed matter physics. The recent discovery of superconductivity in infinite-layer nickelates, as analog to high-$T_c$ cuprates, has opened a new route to tackle this challenge. By growing 8 nm $Pr_{0.8}Sr_{0.2}NiO_2$ films on the $(LaAlO_3)_{0.3}(Sr_2AlTaO_6)_{0.7}$ substrate, we successfully raise the transition temperature $T_c$ from 9 K in the widely studied $SrTiO_3$-substrated nickelates into 15 K. By combining x-ray absorption spectroscopy with the first-principles and many-body simulations, we find a positive correlation between $T_c$ and the pre-edge peak intensity, which can be attributed to the hybridization between Ni and O orbitals induced by the strain. Our result suggests that structural engineering can further enhance unconventional superconductivity, and the charge-transfer property plays a crucial role in the pairing strength.**

The discovery of the high-$T_c$ superconducting cuprates is one of the greatest surprises in quantum materials. Due to the complexity stemming from multiple intertwined orders, the pairing mechanism of cuprates remains an enigma [1]. Many theories have proposed that the unconventional superconductivity is likely associated with the strong correlation nature of the *d*-orbital electrons in transition-metal oxides (TMOs) [2], which can be characterized by the Mott insulating parent state at the filling of one hole per unit cell. This picture has motivated the study of superconductivity in several TMOs and, more recently, magic-angle twisted bilayer graphene [3]. However, the anticipated analogy materials, such as iridates and ruthenates, do not show high $T_c$ as cuprates. Theoretically, it remains a question whether the single-band Hubbard model, as a prototype of correlated electron systems, indeed gives long-range ordered superconductivity [4, 5].

The recent discovery of superconducting nickelates provides a new opportunity to shed light on the puzzle [6–11]. Due to nickelates' structural and chemical similarities with cuprates, a comparative study of these two materials may help establish the pairing mechanism and guide us to design materials with higher $T_c$ [12–23]. Experiments have shown infinite-layer nickelates exhibiting similar magnetic excitations with cuprates [24]. However, nickelates show much lower $T_c$. A possible reason is that the nickelates $RNiO_2$ (R = La, Pr, and Nd) are Mott-Hubbard insulators [18, 22, 25, 26], while cuprates are charge-transfer insulators. The charge-transfer property of the latter is characterized by a prominent pre-peak in the oxygen K-edge x-ray absorption spectrum (XAS), usually referred to as the Zhang-Rice peak [27]. Given that the charge-transfer gap was also found to anti-correlate with $T_c$ in various cuprate compounds [28, 29], it is conceivable that the oxygen-orbital components of dopants may play a crucial role in pairing. To test this postulate, a critical step is to manipulate the charge-transfer extent of nickelates and demonstrate the possibility of enhancing superconductivity [30–32].

Here, we report a striking enhancement of $T_c$ by growing the $Pr_{0.8}Sr_{0.2}NiO_2$ film on a compressive substrate $(LaAlO_3)_{0.3}(Sr_2AlTaO_6)_{0.7}$ (LSAT). Combining first-principles and many-body simulations, we reveal that the in-plane compressive strain strengthens the $p - d$ hybridization between oxygen and nickel orbitals. Thus, a considerable amount of doped carriers distributes into oxygen orbitals, increasing the pre-peak in the oxygen K-edge XAS. This comparison reflects the significant role of the charge-transfer nature in superconductivity and motivates an efficient approach to further enhance superconductivity in TMOs.

## SUBSTRATE-INDUCED STRAIN AND ENHANCED $T_c$

As illustrated in Figs. 1**a**, **b**, we prepared $Pr_{0.8}Sr_{0.2}NiO_2$ films of equal thickness 8 nm on two different substrates, $SrTiO_3$ and LSAT, using pulsed laser deposition (PLD) followed by a $CaH_2$-assisted chemical reduction method. Figs. 1**c**, **d** show the high-angle annular dark-field (HAADF) scanning transmission electron microscopy (STEM) images of the two superconducting $Pr_{0.8}Sr_{0.2}NiO_2$ films on the two substrates, respectively. The contrast difference of diffraction spots reveals the uniform growth of the films with an ordered structure of coherent infinite layer, displaying no evidence of minority phases. The well-defined infinite-layer phase of the $Pr_{0.8}Sr_{0.2}NiO_2$ films is further disclosed in the annular bright-field (ABF) STEM images in Figs. 1**e**, **f**. The clear and sharp interfaces between the films and the substrates confirm that the former are epitaxially strained on the latter. Besides, the reciprocal space maps (RSM) around the (103) reflection demonstrate that the films are strained to the substrates (see Figs. S1**e, f**). Based on both measurements, the in-plane lattice constants of the nickelate films match those of the substrates, *i.e.*, a = 3.868 Å, yielding -1% compressive in-plane strain in the films grown on LSAT compared to $SrTiO_3$ [10].

With the nickelate films on distinct substrates, we first examine their transport properties using a Quantum Design Physical Property Measurement System (PPMS) with a standard four-probe configuration. Figure 2**a** shows the temperature-dependent resistivity of these two $Pr_{0.8}Sr_{0.2}NiO_2$ films on $SrTiO_3$ and LSAT substrates, respectively. Both films exhibit a sudden decrease of resistivity, characteristic of the superconducting transition. Here, we employ the onset temperature to quantify the superconductivity critical temperature $T_c$, resulting in $T_c^{(onset)}$ = 9 K and 15 K (highlighted with arrows) for samples grown on $SrTiO_3$ and LSAT, respectively. To further confirm the superconducting behavior, we investigate the transport properties with a magnetic field (up to 9.0-tesla) applied perpendicular to the ab-plane of the film. As shown in Figs. 2**c**, **d**, the field dependence exhibits similar behaviors: In the normal state, the films display negligible variation, whereas, below the superconducting transition, the conductance is suppressed significantly with an increasing magnetic field.

We summarize the critical temperatures of various nickelate experiments by the gray area in Fig. 2**b** to compare with previous experiments. Considering the sensitivity to the thickness, we further mark the $T_c$ varying from 7 K to 12 K for existing $Pr_{0.8}Sr_{0.2}NiO_2$ studies [9]. The $T_c^{(onset)} \approx 9$ K (denoted by the red diamond) of our 8 nm-thick uncapped $Pr_{0.8}Sr_{0.2}NiO_2$ film grown on $SrTiO_3$ is lower than $T_c^{(onset)} \approx 12$ K from the previous studies on the film with equal thickness. Interestingly, when substrated by LSAT, the same uncapped 8 nm-thick $Pr_{0.8}Sr_{0.2}NiO_2$ film exhibits a considerably higher $T_c^{(onset)} \approx 15$ K, beyond the record of any (capped and uncapped) STO-substrated samples for the same doping. The uncapped films on both substrates in our study show a relatively more significant broadening of superconducting transition, which is probably due to the detailed topotactic reduction and common in various superconducting nickelate films [11, 33]. Nevertheless, the zero resistance develops at 2.0 K and 2.5 K in zero-field for the films on $SrTiO_3$ and LSAT, respectively. We emphasize that the observation of $T_c^{(onset)}$ enhancement for equally thick films on LSAT compared to $SrTiO_3$ is robust in our comprehensive optimal procedure. Since the interfacial effects should be irrelevant for 8 nm-thick nickelates, we attributed this dramatic $T_c^{(onset)}$ enhancement in the comparative study to the crystal structure of $Pr_{0.8}Sr_{0.2}NiO_2$, which is compressively strained by the LSAT substrate.

# PRE-EDGE PEAK IN X-RAY ABSORPTION SPECTRA

To unveil the connection between the strain and the enhancement of $T_c$ in $Pr_{0.8}Sr_{0.2}NiO_2$/LSAT, we further examine their electronic structure using XAS. We select the O K-edge to reveal the nature of dopant carriers, since its characteristic peaks have been regarded as a fingerprint in cuprates [27]. Similar to other recent studies, we find that the K-edge XAS of $Pr_{0.8}Sr_{0.2}NiO_2$/SrTiO$_3$ does not exhibit any pre-edge peak [22, 34], suggesting that the parent compound of superconducting nickelates ($Pr_{0.8}Sr_{0.2}NiO_2$/SrTiO$_3$) fall into the Mott-Hubbard insulator regime. This property is in sharp contrast to cuprates, where an extra pre-edge mobile carrier peak (MCP) arises with hole doping and carries intensity comparable with the doping concentration. We illustrate the difference in XAS features via the partial density of states (PDOS) distribution sketched in Fig. 3**c**. As charge-transfer insulators, cuprates have a relatively strong on-site Coulomb interaction $U_d$ (primarily for the Cu 3$d$ electrons) compared to the charge-transfer energy $\Delta_h$ in the hole language. Therefore, a doped hole primarily occupies the oxygen 2$p$ orbital, which hybridizes with the Cu 3$d$ orbital as the Zhang-Rice singlet band. XAS selectively probes the unoccupied PDOS on oxygen, resulting in the coexistence of the pre-edge MCP and the UHB peaks. Nevertheless, the $\Delta_h$ is slightly larger than the $U_d$ in nickelates, leading to tiny, if any, dopant holes on oxygen orbitals and, accordingly, a single absorption peak. This difference between charge-transfer and Mott-Hubbard insulators is exact only in the atomic limit, while the presence of orbital hybridization mixes them (see the Supplementary Information for detailed discussions).

Unlike the STO-substrated nickelates, the XAS of our $Pr_{0.8}Sr_{0.2}NiO_2$/LSAT sample exhibits an evident pre-peak to the O K main edge at 530.2 eV, reminiscent of the pre-peaks in cuprates [27].

The energy position of 530.2 eV for the pre-peak is nearly equal to the observed hybridization peak in $La_4Ni_3O_8$, which belongs to another family of nickelate materials ($R_{n+1}Ni_nO_{2n+2}$) [35]. Therefore, we can attribute this pre-edge peak in $Pr_{0.8}Sr_{0.2}NiO_2$/LSAT to the MCP. Since the charge-transfer nature appears to be one of the most significant differences between cuprates and nickelates, the emergence of the pre-edge XAS peak in LSAT-substrated nickelates provides insight for its enhanced $T_c$.

## NUMERICAL EXPLANATION OF THE ELECTRONIC STRUCTURE'S CHANGE

To reveal the impact of the strained lattice on nickelates' electronic structure, we simulate the doped carriers and spectral evolution upon straining using density functional theory (DFT) and the extracted many-body model. We first calculate the electronic structure of the pristine and -1% strained $PrNiO_2$ using PBE functional (see Methods). As shown in Fig. 4, the simulated PDOS reflects that HOMO and LUMO orbitals primarily consist of Ni $3d_{x2-y2}$ orbitals. As pointed out in previous studies [22], the rare-earth band also resides at low energy, leading to a "self-doping" effect. To address the experimental features observed in the oxygen K-edge XAS, we focus on the oxygen components of the electronic wavefunction. By projecting the simulated Kohn-Sham orbitals to single-particle Wannier orbitals, we extracted two oxygen orbitals ($2p_x$ and $2p_y$) which hybridize with the active Ni $3d_{x2-y2}$ orbital [see the inset of Fig. 4a]. We employ these three orbitals to construct a tight-binding model, the first four terms in Eq. (1) in the Methods section, where the hopping parameters and site energies are extracted from the Kohn-Sham Wannier orbitals. For the pristine $PrNiO_2$, we obtain $t_{pd} = 1.36$ eV, $t_{pp} = 0.57$ eV, and $E_d - E_p = 3.90$ eV, consistent with previous simulations [22] for the -1% strained $PrNiO_2$, $t_{pd}$ are evidently increased into 1.41 eV, while the changes of other parameters are less than 2%.

We then involve the strong Coulomb interactions, doping, and x-ray core-hole interactions on top of the DFT-extracted model to simulate the XAS spectra and analyze correlation effects. Based on the estimations from previous studies, we set the Hubbard-type Coulomb interactions on the Ni $3d$ and O $2p$ orbitals as $U_d = 6$ eV and $U_p = 2$ eV, respectively. We also set the core-hole attractive interaction $U_c = 3$ eV to match the experimental separation between the pre-edge and main absorption peaks. The XAS spectral intensity is simulated using Eq. (2) in the Methods section. As shown in Fig. 4**b**, we manipulate the strength of $t_{pd}$ and fix other parameters, mimicking the primary impact on the electronic structure induced by the strain effect. Considering possible ambiguities of band parameters extracted from minimal Wannier-folded orbitals, we examine the spectral properties for a wide range of $t_{pd}$ values, from 1.0 to 1.5 eV. The simulated XAS results suggest that increased $t_{pd}$, induced by the compressive strain effect, enhances the pre-edge peak, consistent with our experimental observations in Figs. 3**a**, **b**. As illustrated in Fig. 3**c**, although the strain does not switch the relative size of $U_d$ to $\Delta_h$ in the atomic representation, the increased $t_{pd}$ mixes oxygen and nickel PDOS at the top of the LHB. Therefore, a larger portion of the doped holes resides in oxygens, resulting in the rise of the pre-edge XAS peak.

To further demonstrate the aforementioned mechanism, we calculate the average hole concentrations ⟨$n_p$⟩ on the oxygen $2p_{x/y}$ orbitals. By comparing the undoped and doped XAS results, it is clear that the pre-edge peak reflects the majority of dopant carrier concentrations in oxygens. Therefore, we evaluate the pre-peak intensity and ⟨$n_p$⟩ for a wide range of charge-transfer energy $\Delta_h$ and $t_{pd}$ of the three-band Hubbard model. As shown in Figs. 4**c**, **d**, the parameter dependences of these two quantities are consistent in the entire $\Delta_h - t_{pd}$ parameter space. We also highlight the DFT-extracted model parameters for the pristine and strained nickelates to guide the eye. In this context, the pre-edge peak intensity rises by 2.8%, and the oxygen-hole concentration increases by 3.1%. While the simplicity of the three-band model and the accuracy of the DFT simulated parameters may affect the exact values, the correlation between the spectral and carrier properties qualitatively reflects that the nickelates in LSAT substrate distribute more dopant carriers into oxygen orbitals.

## DISCUSSION AND OUTLOOK

The fact that strained PSNO displays a much higher $T_c$ not only provides an efficient route to control, and especially enhance, the superconductivity in nickelates, but also gives significant insights into unconventional superconductivity in TMOs. Since the rise of $T_c$ is accompanied by the pre-edge peak in oxygen K-edge XAS, we can conclude that the TM-oxygen hybridization plays a crucial role in the pairing mechanism. A natural explanation is the enhanced spin fluctuations due to O-Ni orbital hybridizations, similar to the recently observed strained-enhanced magnons in monolayer cuprates [36]. According to our first-principles extracted three-band models, we simulate the dynamical spin structure factors of the pristine and strained PNO. The top of magnon excitations rises from 178 meV into 192 meV with the -1% compressive strain (see the Supplementary Information). While its impact on $T_c$ remains to be further addressed with unbiased large-scale simulations [37], we may qualitatively argue that the pronounced increase of spin-spin coupling and the increase of hole hopping would increase $T_c$. Large-momentum spin fluctuations strengthen the $d$-wave pairing instability and have been popular candidates for the pairing glue of unconventional superconductivity.

The fact that a 1% strain induces a 7.8% increase of spin excitations indicates the outsized impact of the lattice structure. Thus, our result may also help to examine the possible contribution of electron-phonon to unconventional superconductivity, as suggested by recent studies in cuprates[38–43]. As a key difference from nickelates and other TMOs [26], the unique charge-transfer insulating parent compounds of cuprates cause doped holes primarily to reside on oxygens, whose vibrations constitute the majority of phonon modes. The present experimental and simulation results demonstrate that the strain-enhanced hybridization between O and Ni orbitals also leads to the similar charge transfer of doped carriers, which may further help pairing.

It is still challenging to examine if the magnetic or phonon-assisted mechanism, or their interplay, may account for the dramatic (more than 60%) enhancement of $T_c$ in $Pr_{0.8}Sr_{0.2}NiO_2$/LSAT. The quantitative assessment of their contributions may resolve the long-standing puzzle of the pairing mechanism in cuprates and motivates future experiments. For example, resonant inelastic x-ray scattering (RIXS) has characterized both the magnon and phonon dispersions in STO-substrated $Nd_{1-x}Sr_xNiO_2$ [24]. Comparative RIXS studies of nickelates on these two substrates may provide more insight into individual contributions from these excitations. Thus, our results, together with the further investigations of pairing mechanisms, should stimulate systematic strain-engineering of unconventional superconductivity.

## METHODS

### A. Sample Preparation and Experimental Characterization

The perovskite precursor $Pr_{0.8}Sr_{0.2}NiO_3$ films were prepared by using pulsed laser deposition

(PLD). The corresponding infinite-layer phase was acquired by the soft-chemistry reduction method. The superconducting transition temperature was confirmed by transport measurements using a Quantum Design Physical Property Measurement System (PPMS) with a standard four-probe configuration. Samples for the cross-sectional scanning transmission electron microscopy (STEM) were prepared by focused ion beam (FIB, Helios 600i) techniques. The high-angle annular dark field (HAADF) and annular bright field (ABF) STEM images were acquired on the ARM-200F (JEOL, Japan) operated at 200 kV with a CEOS Cs corrector (CEOS GmbH, Germany). The x-ray absorption spectroscopy (XAS) measurements were performed at beamline 29-ID IEX at the Advanced Photon Source, Argonne National Laboratory. The beamline uses an electromagnetic undulator providing a soft x-ray from 250 eV to 3000 eV. The spectra were normalized by the incident x-ray intensity (I0) using a drain current from a gold mesh upstream of the sample. All spectra were collected at 30K.

## B. Density Functional Theory Calculations

The electronic band structure of $PrNiO_2$ is calculated using the Quantum Espresso package [44] with Perdew-Burke-Ernzerhof (PBE) [45] exchange-correlation functional, where the pseudopotential is based on the projected augmented wave (PAW) method [46]. We adopt the Monkhorst-Pack sampling [47] with a $\Gamma$-centered k-mesh of $13 \times 13 \times 15$ and a plane-wave cutoff energy of 40 rydberg is used to expand the wave function. The convergence criteria of structure relaxation and self-consistent field are set to $10^{-4}$ rydberg/bohr and $10^{-7}$ rydberg, respectively. For the -1 % strained structure, we then change the lengths of the a- and b-axis directly while the c-axis remains unchanged.

To extract the site energies and hopping parameters, we employ the Wannier90 package [48] to fit our DFT results and construct the tight-binding models, where a total number of 16 Wannier functions: five Pr's *d* orbitals, five Ni's *d* orbitals, and six O's *p* orbitals, are considered. In our calculations, the disentanglement procedure is employed with a disentanglement window from -10 to +0.8 eV.

## C. Exact Diagonalization Simulation for XAS Spectra

We employ the three-band Hubbard model extracted from the DFT Wannier orbitals as a minimal description of the charge-transfer physics in PSNO. The Hamiltonian reads as:

$$H_v = -\sum_{\langle i,j\rangle\alpha\sigma} t_{pd}^{ij}(d_{i\sigma}^+ p_{j\alpha\sigma} + h.c.)$$
$$-\sum_{\langle j,j'\rangle\sigma} t_{pp}^{jj'}(p_{jx\sigma}^+ p_{j'y\sigma} + h.c.) + \sum_{i\sigma} E_d n_{i\sigma}^d + \sum_{j\alpha\sigma} E_p n_{j\alpha\sigma}^p$$
$$+\sum_i U_d n_{i\uparrow}^d n_{i\downarrow}^d + \sum_{j\alpha} U_p n_{j\alpha\uparrow}^p n_{j\alpha\downarrow}^p, \quad (1)$$

Here, $d_{i\sigma}^+(p_{j\alpha\sigma}^+)$ creates an electron with spin $\sigma$ at a Ni site $i$ (O site $j$), and $n_{i\sigma}^d(n_{j\sigma}^p)$ is the Ni (O) electron number. The flavor $\alpha = \{x, y\}$ labels the $2p_{x,y}$ orbitals. The $t_{pd}$ and $t_{pp}$ determine the nearest-neighbor hopping amplitudes, and $E_d$ ($E_p$) is the electronic site energy. These single-particle parameters are extracted from the Wannier orbitals of the DFT simulation. The last two terms are the on-site Hubbard interactions, whose strengths are set as *ad hoc* parameters $U_d = 6$ eV, $U_p = 2$ eV, comparable with previous studies [22, 26]. The charge-transfer energy in the hole representation is defined as $\Delta_h = E_d - E_p + U_d - U_p$. At the pristine PSNO, the DFT simulation gives $\Delta_h = 8$ eV, $|t_{pd}| = 1.36$ eV, $|t_{pp}| = 0.57$ eV, while in the -1% strained PSNO, the extracted parameters of $|t_{pd}|$ and $|t_{pp}|$ are =1.41 eV and 0.58 eV. We perform exact diagonalization calculations of the XAS spectrum of the three-orbital Hubbard model on an 8-site cluster for two different doping levels: 0% (half-filling) and 12.5% (underdoped).

With the many-body ground state $|G\rangle$, the zero-temperature XAS spectrum is

$$I_{XAS}(\omega,T) = -\frac{1}{\pi}\text{Im}\langle G|D^+ \frac{1}{\omega+E_G-H+i\delta}D|G\rangle, \tag{2}$$

where $E_G$ is the ground-state energy and $\delta$ is the broadening corresponding to the inverse lifetime of the intermediate state. For the convenience of extracting the integrated peak intensity, we employed a small broadening $\delta = 0.1$ eV in the simulation, where realistic values should be much larger. The dipole operators $D$ define the transformation between the oxygen $1s$ (core-level) and the $2p$ orbitals (valence level) selected by the x-ray K-edge edge

$$D = \sum_{\alpha\sigma} p^+_{j\alpha\sigma}c_{j\sigma}, \tag{3}$$

We ignore the matrix elements since we do not compare absolute intensities among different edges. In addition to the valence-band tight-binding model $H_v$, the intermediate state in Eq. (2) contains an extra core hole and its attractive interaction

$$H_c = \sum_{i\sigma}E_{edge}(1-n^c_{i\sigma}) - U_c\sum_{i\sigma\sigma'}n^p_{i\sigma}(1-n^c_{i\sigma'}), \tag{4}$$

where $E_{\text{edge}}$ is the edge energy $E_{\text{edge}} = 530$ eV and $U_\text{c} = 3$ eV the core-hole potential.

**Data availability**

The data will be released upon final publication.

**ACKNOWLEDGMENTS**

We thank Zhuoyu Chen, Mingda Li, Krzysztof Wohlfeld for helpful discussions and John Freeland for experimental assistance on XAS. The experimental part of this work was supported by the National Natural Science Foundation of China (Grant No. 12074411) and (Grant No. 11888101), the National Key Research and Development Program of China (Grant No. 2016YFA0300300 and 2017YFA0302900), the Strategic Priority Research Program (B) of the Chinese Academy of Sciences (Grant No. XDB25000000) and the Research Program of Beijing Academy of Quantum Information Sciences (Grant No. Y18G06). The first-principles and model-based spectral simulations used resources of the National Energy Research Scientific Computing Center (NERSC), a U.S. Department of Energy Office of Science User Facility operated under Contract No. DE-AC02-05CH11231. The XAS measurement performed at the Advanced Photon Source was supported by the U.S. Department of Energy, Office of Science, and Office of Basic Energy Sciences under Contract No. DE-AC02-06CH11357.

**AUTHOR CONTRIBUTIONS**

X. L. R. and Q. G. prepared and characterized the film samples. X. L. R., Q. G, and H. L. L. performed the transport and STEM measurements. W. C. C., J. H., and Y. W. performed the theoretical and numerical calculations. J. L., J. J. S., F. R., J. L. M, R. C. conducted the XAS measurements. X. L. R., Z. H. Z., and X. J. Z. analyzed the data. Y. W, T. X., and J. P. H., and F. C. Z. provided theoretical understanding. X. L. R., Y. W., and Z. H. Z. wrote the manuscript with input from all authors. X. J. Z. and Z. H. Z. conceived and directed the project. The authors declare no competing financial interests.

**DECLARE OF INTEREST**

The authors declare no competing interests.

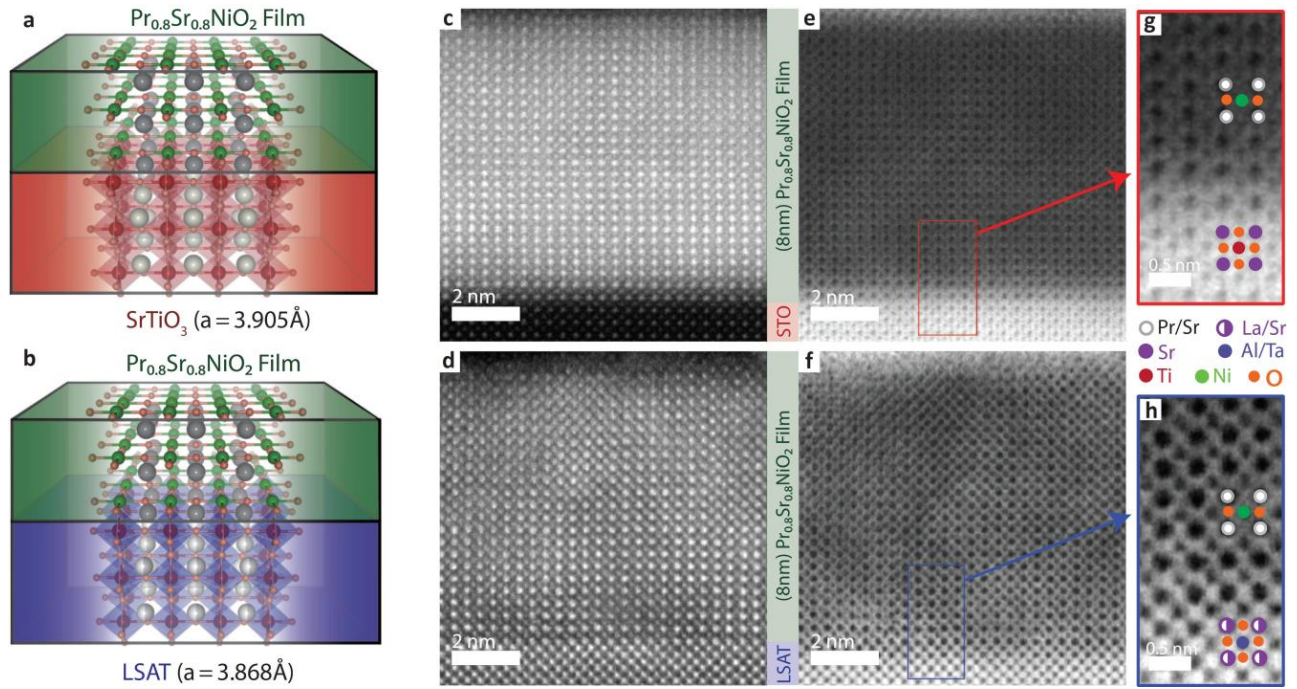

**Fig. 1. Cross-sectional scanning transmission electron microscopy (STEM) images of two types of superconducting $Pr_{0.8}Sr_{0.2}NiO_2$ films. a**, **b**, The schematic illustrations of heterostructures of $Pr_{0.8}Sr_{0.2}NiO_2$ films on two different substrates $SrTiO_3$ and LSAT, respectively. **c**, **d**, The high-angle annular dark field (HAADF) STEM images of films on $SrTiO_3$ and LSAT substrates, respectively. **e**, **f**, The annular bright field (ABF) STEM images of film on $SrTiO_3$ and LSAT substrates, respectively. **g**, **h**, The enlarged views of the areas inside the rectangles in **e** and **f**, respectively.

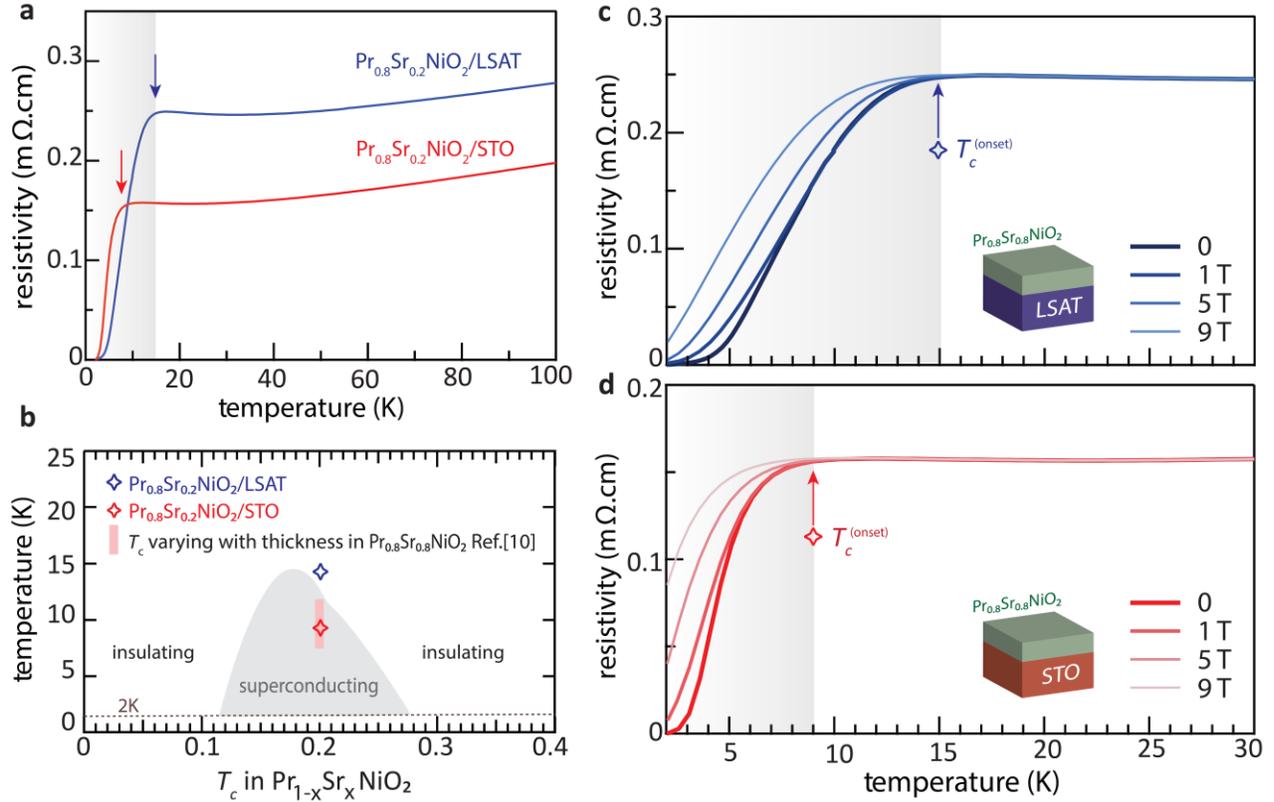

**Fig. 2. Transport properties of superconducting Pr$_{0.8}$Sr$_{0.2}$NiO$_2$ films on two different substrates. a**, Temperature-dependent resistivity of the Pr$_{0.8}$Sr$_{0.2}$NiO$_2$ on SrTiO$_3$ and LSAT substrates, respectively. **b**, The phase diagram of Pr$_{1-x}$Sr$_x$NiO$_2$ films adapted from Ref. [10]. The $T_c$ varies from 7 K to 12 K with thickness for Pr$_{0.8}$Sr$_{0.2}$NiO$_2$ films adapted from Ref. [9] (pink bar). The $T_c^{(onset)} \approx 9$ K (red diamond) and $T_c^{(onset)} \approx 15$ K (blue diamond) of our 8 nm-thick uncapped Pr$_{0.8}$Sr$_{0.2}$NiO$_2$/SrTiO$_3$ and Pr$_{0.8}$Sr$_{0.2}$NiO$_2$/LSAT. **c, d**, Magnetic-field response of superconducting Pr$_{0.8}$Sr$_{0.2}$NiO$_2$ film on SrTiO$_3$ and LSAT substrates, at a varying magnetic field perpendicular to the a-b plane. The onset of superconducting transition at 9 K for the film on SrTiO$_3$ is smaller than that of the film at 15 K on LSAT substrate. Zero-resistance is obtained at 2.5 K and 2.0 K for the films on LSAT and SrTiO$_3$, respectively.

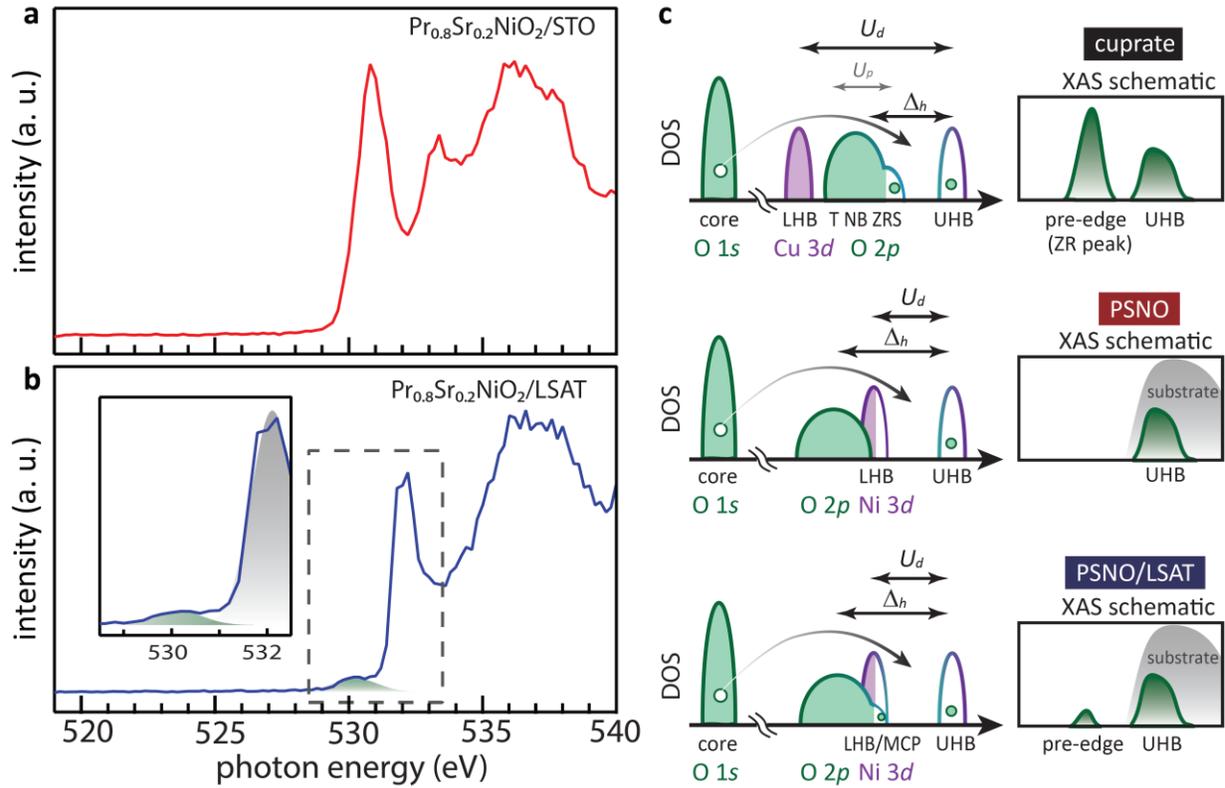

**Fig. 3. Total fluorescence-yield X-ray absorption spectra (XAS) of two superconducting nickelate films. a**, **b**, The XAS at the O K edge of $Pr_{0.8}Sr_{0.2}NiO_2$ thin films with the linear polarization perpendicular to the films' c axis. The inset in **b** shows the fitted spectral features near the pre-edge peak (the regime inside of the dashed square). The mobile carrier peak (MCP) is absent in the film on $SrTiO_3$ substrate of **a**. **c**, Schematic density of states (DOS) of transition-metal oxides (TMOs) and their corresponding XAS features. Upper: a charge-transfer insulator, represented by cuprates, distributes doped holes primarily to oxygen orbitals, resulting in a prominent pre-edge peak in XAS; Middle: a Mott-Hubbard insulator, represented by nickelates on STO substrates, distributes doped holes primarily to transition-metal $3d$ orbitals, resulting in negligible pre-edge peaks; Bottom: strained nickelates exhibit stronger hybridization between $3d$ and $2p$ orbitals, producing a pre-edge peak.

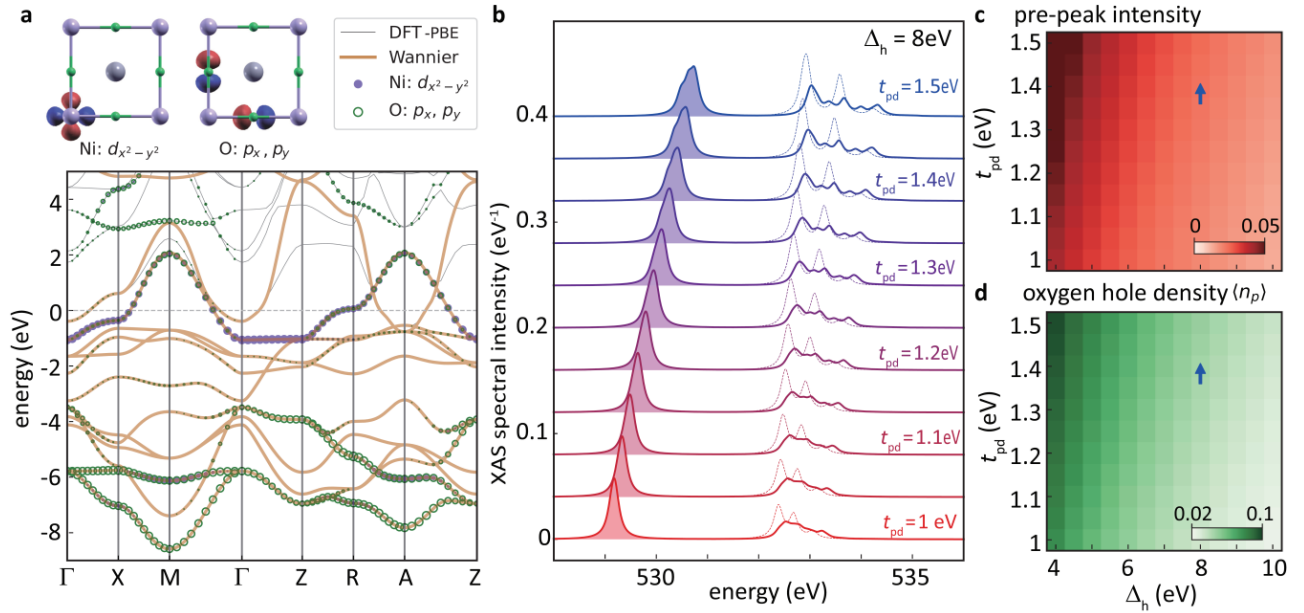

**Fig. 4. Theoretical analysis of strain effects on nickelates and XAS spectroscopies. a**, Band structures of unstrained $PrNiO_2$ from DFT and Wannier tight-binding model. Nickel's $3d_{x^2-y^2}$ and oxygen's $p_x$ and $p_y$ orbitals are shown at the top, where gray, purple, and green spheres represent Pr, Ni, and O, respectively. The site energies of $E_d$ and $E_p$, as well as the hopping parameters $t_{pd}$ (hopping between Ni: $3d_{x^2-y^2}$ and O: $p_x$, $p_y$) and $t_{pp}$ (hopping between O: $p_x$ and $p_y$) in the Wannier tight-binding model. These band parameters are employed in exact diagonalization simulations. **b**, Exact diagonalization X-ray absorption spectra with a $\Delta_h = 8$ eV and a varying $t_{pd}$ from 1.0 eV to 1.5 eV. The dashed and solid lines represent the spectra for 0 and 12.5 % hole-doping, respectively. The pre-peak area (illustrated by the shade) increases with an increasing $t_{pd}$. The pre-peak area **c** and the average hole number per $p_{x,y}$ orbital **d** as a function of $\Delta_h$ and $t_{pd}$ are shown by the intensity plots. Both of them increase with an increasing $t_{pd}$ and a decreasing $\Delta_h$. The blue arrows indicate the change of $t_{pd}$ from 1.36 to 1.41 eV when 1 % compressive epitaxial strain is applied.